# Unification of Gravities with GUTs

*D. Roumelioti* [a,1], *S. Stefas* [a,2], *G. Zoupanos* [a,b,c,d,3]

[a] Physics Department, National Technical University of Athens, Zografou Campus, 157 80, Zografou, Greece

[b] Max-Planck Institut für Physik, Boltzmannstr. 8, 85 748 Garching/Munich, Germany

[c] Universität Hamburg, Luruper Chaussee 149, 22 761 Hamburg, Germany

[d] Deutsches Elektronen-Synchrotron DESY, Notkestraße 85, 22 607, Hamburg, Germany

Within the gauge-theoretic approach of gravity, the gauging of an enlarged symmetry of the tangent space in four dimensions allows gravity to be unified with internal interactions. We study the unification of the Conformal and Noncommutative (Fuzzy) Gravities with Internal Interactions based on the $SO(10)$ GUT.

## 1. Introduction

In addition to the well-known fact that the Standard Model of Particle Physics is described very successfully by gauge theories, it has long been believed that general relativity (GR) can be formulated as a gauge theory [1–7] too. Furthermore, a very interesting suggestion towards unifying gravity as gauge theory with the rest fundamental interactions described as GUTs, has been suggested a few decades ago [8, 9] and revived recently [10–18]. It is based on the observation that although the dimension of the tangent space is usually taken to be equal to the dimension of the corresponding curved manifold, the tangent group of a manifold of dimension $d$ is not necessarily $SO(d)$ [19]. This observation provides us with the very interesting possibility to consider tangent groups of dimension higher than four, in a 4-d spacetime, which in turn could lead to a unification of gravity with internal interactions by gauging these higher-dimensional tangent groups. A very attractive feature of this approach is that most of the tools which have been used in the high dimensional theories with extra physical space dimensions, such as in the coset space dimensional reduction (CSDR) scheme [20–29], can be transferred in the present four-dimensional framework since they have the same tangent group. For instance, there exist constraints which one has to take into account aiming to construct realistic chiral gauge theories describing the internal interactions. An example of such constraint is the Weyl condition that one has to impose in the extra-dimensional gauge theory in order to obtain a chiral one in 4-d. Similarly to avoid a possible doubling of the spectrum in a constructed 4-d chiral gauge theory one has to impose in addition to the Weyl also the Majorana condition in the higher-dimensional one [22, 25].

Along the lines described above, a unification of Conformal Gravity (CG) and internal interactions has been constructed recently [18]. This approach was subsequently extended to the unification of 4-d Gravity on a covariant noncommutative (fuzzy) space, i.e. Fuzzy gravity (FG) with internal interactions [30]. Here, we present the basic features of these constructions.

---

[1] E-mail: danai_roumelioti@mail.ntua.gr
[2] E-mail: dstefas@mail.ntua.gr
[3] E-mail: george.zoupanos@cern.ch

## 2. Conformal Gravity

The Einstein Gravity (EG) has been treated as the gauge theory of the Poincaré group [2] but much insight and elegance was gained by considering instead the gauging of the de Sitter (dS) and the Anti-de Sitter (AdS) groups, $SO(1,4)$ and $SO(2,3)$ respectively. Both groups contain the same number of generators, i.e. 10, as the Poincaré group and can be spontaneously broken by a scalar field to the Lorentz group, $SO(1,3)$ [6, 7, 18, 31]. Note, in addition, that the Poincaré, the dS and the AdS groups are all subgroups of the conformal group $SO(2,4)$, which has 15 generators. In ref [32] the gauge theory formalism of Gravity was extended to the conformal group constructing in this way the CG. The breaking of CG to EG or to Weyl's scale invariant theory of gravity was done by the imposition of constraints (see e.g. [32]). However in ref [18], for the first time, the breaking of the conformal gauge group was done spontaneously introducing a scalar field in the action and using the Lagrange multiplier procedure.

The Spontaneous Symmetry Breaking (SSB) of CG, which is based, as already mentioned on the $SO(2,4)$ gauge group, whose algebra is isomorphic to those of $SU(4)$ and $SO(6)$, can be done in two ways. For convenience we work with Euclidean signature. Then one way to obtain an SSB towards EG is to introduce a scalar in the vector representation (rep) of $SO(6)$, 6, which takes vev in the $\langle 1 \rangle$ component [33, 34] of the 6 according to the branching rules of reps of $SO(6)$ to its maximal subgroup $SO(5)$ [18], i.e.

$$SO(6) \supset SO(5) \tag{1}$$
$$6 = 1 + 5 \,. \tag{2}$$

Then the $SO(5)$, being isomorphic to $SO(2,3)$, can break spontaneously further to $SO(1,3)$, when a scalar in the 5 rep takes a vev in the $\langle 1,1 \rangle$ component according to the branching rules:

$$SO(5) \supset SU(2) \times SU(2) \tag{3}$$
$$5 = (1,1) + (2,2) \,, \tag{4}$$

where the algebra of $SU(2) \times SU(2)$ is isomorphic to those of $SO(4)$ and $SO(1,3)$. Therefore in order to realize the above breakings from $SO(2,4)$ to $SO(1,3)$ one has to introduce two scalar fields, belonging in the vector rep, 6 of $SO(2,4)$ (for details see ref [18]).

In addition to the above way of breaking the $SO(2,4)$ to $SO(1,3)$, we can use a more direct way to achieve the same breaking in one step using a scalar belonging in the 2nd rank antisymmetric rep of $SO(6)$, i.e. the rep 15. In fact, this breaking can lead either to the four-dimensional EG or to Weyl Gravity (WG) as we will be discussed later in this section.

The gauge group $SO(2,4)$, as mentioned previously, has 15 generators. These generators in four-dimensional notation can be represented by the six Lorentz transformations, $M_{ab}$, four translations, $P_a$, four special conformal transformations (conformal boosts), $K_a$, and the dilatation D. The gauge connection, $A_\mu$, as an element of the $SO(2,4)$ algebra, can be expanded in terms of the generators as

$$A_\mu = \frac{1}{2}\omega_\mu{}^{ab} M_{ab} + e_\mu{}^a P_a + b_\mu{}^a K_a + \tilde{a}_\mu D \,, \tag{5}$$

where a gauge field has been introduced for each generator. The gauge field related to the translations is identified as the vierbein, while the one of the Lorentz transformations is identified as the spin connection. Then the field strength tensor is of the form

$$F_{\mu\nu} = \frac{1}{2}R_{\mu\nu}{}^{ab}M_{ab} + \tilde{R}_{\mu\nu}{}^{a}P_a + R_{\mu\nu}{}^{a}K_a + R_{\mu\nu}D,\qquad(6)$$

and its components can be found in ref [18].

The action is chosen to be parity conserving and quadratic in terms of the field strength tensor (6), in which we have introduced a scalar $\phi$ that belongs in the 2nd rank antisymmetric rep, 15, of $SO(6) \sim SO(2,4)$ along with a dimensionfull parameter, m:

$$S_{SO(2,4)} = a_{CG}\int d^4x\left[\text{tr}\,\epsilon^{\mu\nu\rho\sigma}m\phi F_{\mu\nu}F_{\rho\sigma} + \left(\phi^2 - m^{-2}\mathbb{1}_4\right)\right],\qquad(7)$$

where the trace is defined as $tr \to \epsilon_{abcd}[\text{Generators}]^{abcd}$.

The scalar, expanded on the generators, can be written as:

$$\phi = \phi^{ab}M_{ab} + \tilde{\phi}^a P_a + \phi^a K_a + \tilde{\phi}D.\qquad(8)$$

In accordance with [35], we pick the specific gauge in which $\phi$ is diagonal of the form $\text{diag}(1,1,-1,-1)$. Specifically we choose $\phi$ to be only in the direction of the dilatation generator D:

$$\phi = \phi^0 = \tilde{\phi}D \xrightarrow{\phi^2 = m^{-2}\mathbb{1}_4} \phi = -2m^{-1}D.\qquad(9)$$

In this particular gauge the action reduces to

$$S = -2a_{CG}\int d^4x\,\text{tr}\,\epsilon^{\mu\nu\rho\sigma}F_{\mu\nu}F_{\rho\sigma}D,\qquad(10)$$

and the gauge fields $e$, $b$ and $\tilde{a}$ become scaled as $me$, $mb$ and $m\tilde{a}$ correspondingly. After straightforward calculations, using the expansion of the field strength tensor and the anticommutation relations of the generators, we obtain [18]:

$$S_{\text{SO}(1,3)} = \frac{a_{CG}}{4}\int d^4x\epsilon^{\mu\nu\rho\sigma}\epsilon_{abcd}R_{\mu\nu}^{ab}R_{\rho\sigma}{}^{cd}\qquad(11)$$

with its invariance having obviously been reduced only to Lorentz. Before continuing, we notice that there is no term containing the field $\tilde{a}_\mu$ present in the action. Thus, we may set $\tilde{a}_\mu = 0$. This simplifies the form of the two component field strength tensors related to the P and K generators:

$$\begin{aligned}\tilde{R}_{\mu\nu}{}^a &= mT^{(0)a}_{\mu\nu}(e) - 2m^2\tilde{a}_{[\mu}e_{\nu]}{}^a \longrightarrow mT^{(0)a}_{\mu\nu}(e),\\ R_{\mu\nu}{}^a &= mT^{(0)a}_{\mu\nu}(b) + 2m^2\tilde{a}_{[\mu}b_{\nu]}{}^a \longrightarrow mT^{(0)a}_{\mu\nu}(b).\end{aligned}\qquad(12)$$

The absence of the above field strength tensors in the action, allows us to also set $\tilde{R}_{\mu\nu}{}^a = R_{\mu\nu}{}^a = 0$, and thus to obtain a torsion-free theory. Since $R_{\mu\nu}$ is also absent from the expression of the broken action, it may also be set equal to zero. Then we obtain the following relation among $e$ and $b$:

$$e_\mu{}^a b_{\nu a} - e_\nu{}^a b_{\mu a} = 0.\qquad(13)$$

The above result reinforces one to consider solutions that relate e and b. Here we examine two interesting solutions of relation(13):

**A.** When $\boxed{b_\mu{}^a = ae_\mu{}^a}$, we obtain the Einstein-Hilbert action in the presence of a cosmological constant:

$$S_{\text{SO}(1,3)} = \frac{a_{CG}}{4} \int d^4x \epsilon^{\mu\nu\rho\sigma}\epsilon_{abcd} \left[ R^{(0)ab}_{\mu\nu} R^{(0)cd}_{\rho\sigma} - 16m^2 a R^{(0)ab}_{\mu\nu} e_\rho{}^c e_\sigma{}^d + \right. \tag{14}$$
$$\left. +64m^4 a^2 e_\mu{}^a e_\nu{}^b e_\rho{}^c e_\sigma{}^d \right] .$$

**B.** When $\boxed{b_\mu{}^a = -\frac{1}{4}(R_\mu{}^a - \frac{1}{6}Re_\mu{}^a)}$, we obtain the Weyl action:

$$S = \frac{a_{CG}}{4} \int d^4x \epsilon^{\mu\nu\rho\sigma}\epsilon_{abcd} C_{\mu\nu}{}^{ab} C_{\rho\sigma}{}^{cd} = 2a_{CG} \int d^4x \left( R_{\mu\nu}R^{\nu\mu} - \frac{1}{3}R^2 \right), \tag{15}$$

where $C_{\mu\nu}{}^{ab}$ is the Weyl conformal tensor.

### 3. Noncommutative Gauge Theory of 4D Gravity - Fuzzy Gravity

#### 3.1. The Background Space

Before we move on with the gauge theory of FG, we will first have to establish the background space, on which this theory will be formulated. In refs [36–40] extending the original Snyder's suggestion [41] the authors have considered the group the $SO(1,5)$ and have assigned the 4-d space-time coordinates to elements of its Lie algebra.

More specifically we start with the group $SO(1,5)$, whose generators obey the following Lie algebra:

$$[J_{mn}, J_{rs}] = i \left( \eta_{mr} J_{ns} + \eta_{ns} J_{mr} - \eta_{nr} J_{ms} - \eta_{ms} J_{nr} \right), \tag{16}$$

where $m, n, r, r = 0, \ldots, 5$, and $\eta_{mn} = diag(-1, 1, 1, 1, 1, 1)$. Performing the decompositions $SO(1,5)$ to its maximal subgroups up to $SO(1,3)$, i.e., $SO(1,5) \supset SO(1,4)$ and $SO(1,4) \supset SO(1,3)$, turns the above commutation relation to the following:

$$[J_{ij}, J_{kl}] = i \left( \eta_{ik} J_{jl} + \eta_{jl} J_{ik} - \eta_{jk} J_{il} - \eta_{il} J_{jk} \right), \quad [J_{ij}, J_{k5}] = i \left( \eta_{ik} J_{j5} - \eta_{jk} J_{i5} \right),$$
$$[J_{i5}, J_{j5}] = iJ_{ij}, \quad [J_{ij}, J_{k4}] = i \left( \eta_{ik} J_{j4} - \eta_{jk} J_{i4} \right), \quad [J_{i4}, J_{j4}] = iJ_{ij}, \tag{17}$$
$$[J_{i4}, J_{j5}] = i\eta_{ij} J_{45}, \quad [J_{ij}, J_{45}] = 0, \quad [J_{i4}, J_{45}] = -iJ_{i5}, \quad [J_{i5}, J_{45}] = iJ_{i4}.$$

Next, it is possible to convert the generators to physical quantities by setting $\Theta_{ij} = \hbar J_{ij}$, and $X_i = \lambda J_{i5}$, where $\lambda$ is a natural unit of length. Furthermore, the momenta can be identified as $P_i = \frac{\hbar}{\lambda} J_{i4}$, and $h = J_{45}$. Given these identifications as well as the generators' algebra, we can calculate the commutation relations of the above quantities:

$$[\Theta_{ij}, \Theta_{kl}] = i\hbar \left( \eta_{ik}\Theta_{jl} + \eta_{jl}\Theta_{ik} - \eta_{jk}\Theta_{il} - \eta_{il}\Theta_{jk} \right), \quad [\Theta_{ij}, X_k] = i\hbar \left( \eta_{ik} X_j - \eta_{jk} X_i \right),$$
$$[\Theta_{ij}, P_k] = i\hbar \left( \eta_{ik} P_j - \eta_{jk} P_i \right), \quad [X_i, X_j] = \frac{i\lambda^2}{\hbar}\Theta_{ij}, \quad [P_i, P_j] = \frac{i\hbar}{\lambda^2}\Theta_{ij}, \tag{18}$$
$$[X_i, P_j] = i\hbar\eta_{ij} h, \quad [\Theta_{ij}, h] = 0, \quad [X_i, h] = \frac{i\lambda^2}{\hbar} P_i, \quad [P_i, h] = -\frac{i\hbar}{\lambda^2} X_i.$$

Given the above relations, the following significant results may be drawn. First, since the coordinates as well as the momenta are elements of this Lie algebra, they exhibit a noncommutative behavior, implying that both the space-time and the momentum space become quantized. In addition, it becomes evident that the commutation relation between coordinates and momenta naturally yields a Heisenberg-type uncertainty relation.

### 3.2. Gauge Group and Representation

Starting with the formulation of a gauge theory for gravity in the space mentioned above, we first have to determine the group that will be gauged. Naturally, the group we will choose is the one that describes the symmetries of the theory, in this case, the isometry group of $dS_4$, $SO(1,4)$. Since, as is shown in refs. [38, 39], in noncommutative gauge theories the use of anticommutators of the gauge group generators is inevitable, and since the anticommutators of the generators of the above isometry group do not necessarily yield elements that belong in the same algebra, we have to take into account the closing of the anticommutators of the relevant gauge group generators. In order to achieve that, we have to pick a specific representation of the algebra generators, and subsequently extend the initial gauge group to one with larger symmetry, in which both the commutator and anticommutator algebras close. Following this procedure, we are led to the extension of our initial gauge group $SO(1,4)$ to $SO(2,4) \times U(1)$.

### 3.3. Fuzzy Gravity

Having determined the appropriate gauge group, we are ready to move on with the (noncommutative) gauging procedure on the present, fuzzy, background space.

First, one has to define the covariant coordinate of the theory, which is given by:

$$\mathcal{X}_\mu = X_\mu \otimes \mathbb{1}_4 + A_\mu(X), \tag{19}$$

where $A_\mu$ is the gauge connection, which in turn can be expanded on the gauge group generators as:

$$A_\mu = a_\mu \otimes \mathbb{1}_4 + \omega_\mu{}^{ab} \otimes M_{ab} + e_\mu{}^a \otimes P_a + b_\mu{}^a \otimes K_a + \tilde{a}_\mu \otimes D. \tag{20}$$

Given the above expansion (20), we can now explicitly rewrite the covariant coordinate (19) as:

$$\mathcal{X}_\mu = (X_\mu + a_\mu) \otimes \mathbb{1}_4 + \omega_\mu{}^{ab} \otimes M_{ab} + e_\mu{}^a \otimes P_a + b_\mu{}^a \otimes K_a + \tilde{a}_\mu \otimes D. \tag{21}$$

What is left to be determined is the appropriate, covariant, field strength tensor of the theory, which for the noncommutative case is defined as [38, 42]:

$$\hat{F}_{\mu\nu} \equiv [\mathcal{X}_\mu, \mathcal{X}_\nu] - \kappa^2 \hat{\Theta}_{\mu\nu}, \tag{22}$$

where $\hat{\Theta}_{\mu\nu} \equiv \Theta_{\mu\nu} + \mathcal{B}_{\mu\nu}$, and $\mathcal{B}_{\mu\nu}$ a 2-form field taking care of the transformation of $\Theta$, promoting it to its covariant form. Since $\hat{F}_{\mu\nu}$ is an element of the gauge algebra it can also be expanded on the algebra's generators as

$$\hat{F}_{\mu\nu} = R_{\mu\nu} \otimes \mathbb{1}_4 + \frac{1}{2} R_{\mu\nu}{}^{ab} \otimes M_{ab} + \tilde{R}_{\mu\nu}{}^a \otimes P_a + R_{\mu\nu}{}^a \otimes K_a + \tilde{R}_{\mu\nu} \otimes D. \tag{23}$$

The SSB goes along the same lines as the one described in the case of Conformal Gravity, i.e. we introduce a scalar field, $\Phi(X)$, belonging the 2nd rank antisymmetric rep of $SO(2,4)$, in the action and fix it in the gauge that leads to the Lorentz group (see [18, 38, 39]). This scalar field must be charged under the $U(1)$ gauge symmetry in order to break it. Introducing the scalar field, the action takes the form:

$$\mathcal{S} = \mathrm{Trtr}\left[\lambda \Phi(X) \varepsilon^{\mu\nu\rho\sigma} \hat{F}_{\mu\nu} \hat{F}_{\rho\sigma} + \eta \left(\Phi(X)^2 - \lambda^{-2} \mathbb{1}_N \otimes \mathbb{1}_4\right)\right], \tag{24}$$

where $\eta$ is a Lagrange multiplier, and $\lambda$ is a dimensionfull parameter. The resulting action now bears the remaining $SO(1,3)$ gauge symmetry, after the spontaneous symmetry breaking. Furthermore, as it can be seen in [39], when the commutative limit of the action is considered, the latter reduces to the Palatini action, which in turn is equivalent to EG in the presence of a cosmological constant term.

## 4. Unification of Conformal and Fuzzy Gravities with Internal Interactions, Fermions and Breakings

In [18], it was suggested that the unification of the CG with internal interactions based on a framework that results in the GUT $SO(10)$ could be achieved using the $SO(2,16)$ as unifying gauge group. As it was emphasized in the Introduction the whole strategy was based on the observation that the dimension of the tangent space is not necessarily equal to the dimension of the corresponding curved manifold [19], [8–14, 16, 17, 30]. An additional fundamental observation [18] is that in the case of $SO(2,16)$ one can impose Weyl and Majorana conditions on fermions [43, 44]. More specifically, using Euclidean signature for simplicity (the implications of using non-compact space are explicitly discussed in [18]), one starts with $SO(18)$ and with the fermions in its spinor representation, 256. Then the spontaneous symmetry breaking of $SO(18)$ leads to its maximal subgroup $SO(6) \times SO(12)$ [18]. Let us recall for convenience the branching rules of the relevant reps [33–35],

$$\begin{aligned}
SO(18) &\supset SO(6) \times SO(12) \\
256 &= (4, \overline{32}) + (\overline{4}, 32) \qquad\qquad \text{spinor} \\
170 &= (1,1) + (6,12) + (20', 1) + (1,77) \quad \text{2nd rank symmetric}
\end{aligned} \qquad (25)$$

The breaking of $SO(18)$ to $SO(6) \times SO(12)$ is done by giving a vev to the $\langle 1,1 \rangle$ component of a scalar in the 170 rep. In turn, given that the Majorana condition can be imposed, due to the non-compactness of the used $SO(2,16) \sim SO(18)$, we are led after the spontaneous symmetry breaking to the $SO(6) \times SO(12)$ gauge theory with fermions in the $(4, \overline{32})$ representation. Then, according to [18], the following spontaneous symmetry breakings can be achieved by using scalars in the appropriate representations

$$SO(6) \to SU(2) \times SU(2), \qquad (26)$$

in the CG sector, and

$$SO(12) \to SO(10) \times [U(1)]_{\text{global}} \qquad (27)$$

in the internal gauge symmetry sector, with fermions in the $16_L(-1)$ under the $SO(10) \times [U(1)]_{\text{global}}$. The other generations are introduced as usual with more chiral fermions in the 256 rep of $SO(18)$. In order to break the CG gauge group we choose scalars in the 2nd rank antisymmetric 15 rep of $SO(6)$, while the internal interactions gauge group $SO(12)$ is broken spontaneously by scalars in the 77 rep. The 15 rep can be drawn from the $SO(18)$ rep 153:

$$153 = (15,1) + (6,12) + (1,66), \qquad (28)$$

while from (25) we see that the 77 rep can result from a 170 rep of the parent group. Thus, in $SO(6) \times SO(12)$ notation, the scalars breaking the two gauge groups belong to $(15,1)$ and $(1,77)$, respectively. According to the above picture we start from some high scale where the $SO(18)$ gauge group breaks, eventually obtaining EG and $SO(10) \times [U(1)]_{\text{global}}$ after several symmetry breakings. From that point, we use the symmetry breaking paths and field content followed in [45, 46], in order to finally arrive at the SM. In particular, the $SO(10)$ group breaks spontaneously into an intermediate group which eventually breaks into the SM gauge group. The intermediate groups are the Pati-Salam (PS) gauge group, $SU(4)_C \times SU(2)_L \times SU(2)_R$, with or without a discrete left-right symmetry, $D$, and the minimal left-right gauge group (LR), $SU(3)_C \times SU(2)_L \times SU(2)_R \times U(1)_{B-L}$, again with or without the discrete left-right symmetry.

We would like to add a comment about the case of FG. As it was explained in [30], when attempting to unify FG with internal interactions, along the lines of Unification of Conformal Gravity with $SO(10)$ [18], the difficulties that in principle one is facing are that fermions should **(a)** be chiral in order to have a chance to survive in low energies and not receive masses as the Planck scale, **(b)** appear in a matrix representation, since the constructed FG is a matrix model. Then it was suggested [30] and given that the Majorana condition can be imposed, a solution satisfying the conditions **(a)** and **(b)** above is the following. We choose to start with the $SO(6) \times SO(12)$ as the initial gauge theory with fermions in the $(4, \overline{32})$ representation satisfying in this way the criteria to obtain chiral fermions in tensorial representation of a fuzzy space. Another important point is that using the gauge-theoretic formulation of gravity to construct the FG one is led to in gauging the $SO(6) \times U(1) \sim SO(2,4) \times U(1)$. Therefore, from this point of view, there exists only a small difference in the low energy analysis as compared to the CG.

## 5. Conclusions

In a previous paper [18], a potentially realistic model was constructed based on the idea that unification of gravity and internal interactions in four dimensions can be achieved by gauging an enlarged tangent Lorentz group. This possibility was based on the observation that the dimension of the tangent space is not necessarily equal to the dimension of the corresponding curved manifold. In [18], due to the very interesting fact that gravitational theories can be described by gauge theories, first was constructed the CG in a gauge theoretic manner by gauging the $SO(2,4)$ group. Of particular interest was the fact that the spontaneous symmetry breaking of the constructed CG could lead, among others, to the EG and the WG. Then it is was possible to unify the CG with internal interactions based on the $SO(10)$ GUT, using the higher-dimensional tangent group $SO(2,16)$. Including fermions and suitably applying the Weyl and Majorana conditions led to a fully unified scheme, which was further examined concerning its behaviour in low energies in ref [46]. A similar analysis was done for the FG case, in ref [30], starting from the $SO(2,4) \times SO(12)$, with fermions in the $(4, \overline{32})$.

## References

[1] R. Utiyama, Phys. Rev. **101**, 1597 (1956).

[2] T. W. B. Kibble, Journal of Mathematical Physics, 212–221 (1961).

[3] S. W. MacDowell and F. Mansouri, Phys. Rev. Lett. **38**, 739 (1977).

[4] E. Ivanov and J. Niederle, in 9th Int. Col. on Group Theor. Meth. in Phys. (1980).

[5] E. Ivanov and J. Niederle, Phys. Rev. D **25**, 988 (1982).

[6] K. S. Stelle and P. C. West, Phys. Rev. D **21**, 1466 (1980).

[7] T. Kibble and K. Stelle, Prog. In Q. F. T., Imperial-TP-84-85-13 (1985).

[8] R. Percacci, Phys. Lett. B **144**, 37 (1984).

[9] R. Percacci, Nucl. Phys. B **353**, 271 (1991).

[10] F. Nesti and R. Percacci, J. Phys. A **41**, 075405 (2008).

[11] F. Nesti and R. Percacci, Phys. Rev. D **81**, 025010 (2010).

[12] A. H. Chamseddine and V. Mukhanov, JHEP **03**, 033 (2010).

[13] A. H. Chamseddine and V. Mukhanov, JHEP **03**, 020 (2016).

[14] K. Krasnov and R. Percacci, Class. Quant. Grav. **35**, 143001 (2018).

[15] G. Manolakos et al., Eur. Phys. J. ST **232**, 3607 (2023).

[16] S. Konitopoulos, D. Roumelioti, and G. Zoupanos, Fortsch. Phys. **2023**, 2300226 (2023).

[17] P. Schupp, K. Anagnostopoulos, and G. Zoupanos, Eur.Phys.J.ST **232**, 1 (2024).

[18] D. Roumelioti, S. Stefas, and G. Zoupanos, Eur. Phys. J. C **84**, 577 (2024).

[19] S. Weinberg, Fifth Workshop on Grand Unification (1984).

[20] P. Forgåcs and N. Manton, Commun. Math. Phys. **72**, 15 (1980).

[21] N. Manton, Nucl. Phys. B **193**, 502 (1981).

[22] G. Chapline and R. Slansky, Nucl. Phys. B **209**, 461 (1982).

[23] D. Lust and G. Zoupanos, Phys. Lett. B **165**, 309 (1985).

[24] Y. A. Kubyshin et al., Vol. 349 (Springer, 1989).

[25] D. Kapetanakis and G. Zoupanos, Phys. Rept. **219**, 4 (1992).

[26] P. Manousselis and G. Zoupanos, JHEP **2004**, 025 (2004).

[27] A. Chatzistavrakidis and G. Zoupanos, JHEP **09**, 077 (2009).

[28] N. Irges and G. Zoupanos, Phys. Lett. B **698**, 146 (2011).

[29] G. Manolakos, G. Patellis, and G. Zoupanos, Phys. Lett. B **813**, 136031 (2021).

[30] D. Roumelioti, S. Stefas, and G. Zoupanos, Fortschr. Phys. **72**, 2400126 (2024).

[31] G. Manolakos, PhD thesis (Natl. Tech. U., Athens, 2019).

[32] M. Kaku, P. Townsend, and P. van Nieuwenhuizen, Phys. Rev. D **17**, 3179 (1978).

[33] R. Slansky, Phys. Rept. **79**, 1 (1981).

[34] R. Feger., T. Kephart, and R. Saskowski, Comput. Phys. Commun. **257**, 107490 (2020).

[35] L.-F. Li, Phys. Rev. D **9**, 1723 (1974).

[36] C. N. Yang, Phys. Rev. **72**, 874 (1947).

[37] J. J. Heckman and H. Verlinde, Nucl. Phys. B **894**, 58–74 (2015).

[38] G. Manolakos, P. Manousselis, and G. Zoupanos, JHEP **2020** (2020).

[39] G. Manolakos, P. Manousselis, and G. Zoupanos, Fortsch. Phys. **69**, 2100085 (2021).

[40] G. Manolakos et al., Universe **8**, 215 (2022).

[41] H. S. Snyder, Phys. Rev. **71**, 38 (1947).

[42] J. Madore, Classical and Quantum Gravity **9**, 69 (1992).

[43] R. D'Auria et al., J. Geom. Phys. **40**, 101 (2001).

[44] J. Figueroa-O'Farrill, School of Mathematics, Universiry of Edinburgh.

[45] A. Djouadi et al., Eur. Phys. J. C **83**, 10.1140/epjc/s10052-023-11696-4 (2023).

[46] G. Patellis, N. Tracas, and G. Zoupanos, 10.48550/arXiv.2412.02786 (2024).